\documentclass[fleqn,usenatbib]{mnras}
\usepackage{newtxtext,newtxmath}
\usepackage[T1]{fontenc}
\DeclareRobustCommand{\VAN}[3]{#2}
\let\VANthebibliography\thebibliography
\def\thebibliography{\DeclareRobustCommand{\VAN}[3]{##3}\VANthebibliography}
\usepackage{graphicx}
\usepackage{amsmath}
\usepackage{hyperref}

\title[The origin of DPGs in MaNGA Survey]{The origin of double-peaked narrow emission-line galaxies in MaNGA Survey}

\author[Z. Zhang et al.]{
Zhiyun Zhang$^{1,2,3}$\href{https://orcid.org/0009-0003-8089-3602}{\includegraphics[scale=0.07]{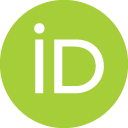}},
Yanmei Chen$^{1,2,3}$\thanks{E-mail: chenym@nju.edu.cn}\href{https://orcid.org/0000-0003-3226-031X}{\includegraphics[scale=0.07]{ORCIDiD_icon128x128.png}},
Shiyin Shen$^{4,5}$\href{https://orcid.org/0000-0002-3073-5871}{\includegraphics[scale=0.07]{ORCIDiD_icon128x128.png}},
Guinevere Kauffmann$^{6}$,
Min Bao$^{1,2,3}$\href{https://orcid.org/0009-0005-9342-9125}{\includegraphics[scale=0.07]{ORCIDiD_icon128x128.png}},\and
\hspace{1mm}Zhijie Zhou$^{1,2,3}$\href{https://orcid.org/0000-0003-1709-6005}{\includegraphics[scale=0.07]{ORCIDiD_icon128x128.png}},
Gaoxiang Jin$^{6}$\href{https://orcid.org/0000-0003-3087-318X}{\includegraphics[scale=0.07]{ORCIDiD_icon128x128.png}}
and Yuren Zhou$^{1,2,3}$\href{https://orcid.org/0000-0001-7785-0626}{\includegraphics[scale=0.07]{ORCIDiD_icon128x128.png}}\\
\\
$^{1}$School of Astronomy and Space Science, Nanjing University, Nanjing 210093, China\\
$^{2}$Key Laboratory of Modern Astronomy and Astrophysics (Nanjing University), Ministry of Education, Nanjing 210093, China\\
$^{3}$Collaborative Innovation Center of Modern Astronomy and Space Exploration, Nanjing 210093, China\\
$^{4}$Shanghai Astronomical Observatory, Chinese Academy of Sciences, 80 Nandan Road, Shanghai 200030, China\\
$^{5}$Key Lab for Astrophysics, Shanghai 200034, China\\
$^{6}$Max Planck Institute for Astrophysics, Karl-Schwarzschild-Str. 1, D-85741 Garching, Germany}

\date{Accepted XXX. Received YYY; in original form ZZZ}
\pubyear{2025}

\begin{document}
\label{firstpage}
\pagerange{\pageref{firstpage}--\pageref{lastpage}}
\maketitle

\begin{abstract}
We select 36 double-peaked narrow emission-line galaxies (DPGs) from 10,010 unique galaxies in MaNGA survey. 
These DPGs show double-peaked Balmer lines and forbidden lines in the spectra.
We use a double Gaussian model to separate the double-peaked profiles of each emission line into blue and red components ($\lambda_\text{blue}$ < $\lambda_\text{red}$), and analyze the spatially resolved kinematics and ionization mechanisms of each component.
We find that in 35 out of 36 DPGs, the flux ratio between the blue and red components varies systematically along the major axes, while it keeps roughly a constant along the minor axes.
The blue and red components of these DPGs exhibit similar distributions in both the value of line-of-sight velocity and the velocity dispersion.
Additionally, 83.3\% DPGs have both blue and red components located in the same ionization region in the ${\hbox{[S\,{\sc ii}]}}$-BPT diagram.
Combining all these observational results, we suggest that the double-peaked emission line profiles in these 35 DPGs primarily originate from rotating discs.
The remaining one galaxy shows clear outflow features.
8 out of 35 DPGs show symmetric line profiles that indicate undisturbed rotating discs, and the other 27 DPGs exhibit asymmetric profiles, suggesting dynamic disturbances in the rotating discs.
Furthermore, we find that 58.3\% DPGs experienced external processes, characterized by tidal features, companion galaxies, as well as gas-star misalignments. This fraction is about twice as much as that of the control sample, suggesting the origin of double-peaked emission line profiles is associated with external processes.
\end{abstract}


\section{Introduction}
The emission-line spectra of galaxies contain a wealth of physical information for the ionized gas, such as kinematics and ionization mechanisms \citep{2019ARA&A..57..511K}.
Typically, emission lines in galaxy spectra are described by a single Gaussian or Lorentzian profile.
Double-peaked emission line profiles were discovered in several
individual objects in early narrow-line region (NLR) studies 
\citep[e.g. Mrk 78, 3C 305 and NGC 4151;][]{1972ApJ...173....7S, 1981ApJ...247..403H, 1984ApJ...281..525H}, suggesting the presence of two distinct gas components with different line-of-sight velocities.

In the past 15 years, systematic searches for kiloparsec-scale dual AGNs have focused on galaxies with double-peaked narrow emission lines.
Using the spatially integrated spectra, \citet{2009ApJ...705L..76W},  \citet{2010ApJ...708..427L}, and \citet{2010ApJ...716..866S} selected double-peaked AGN samples from Data Release 7 of the Sloan Digital Sky Survey \citep[SDSS DR7;][]{2009ApJS..182..543A}, which in total contain 340 AGNs with double-peaked ${\hbox{[O\,{\sc iii}]}}$$\lambda$5007 emission line profiles.
However, in follow-up multi-band photometric and spectroscopic observations of these objects, only a small fraction of them have been confirmed as dual AGNs through high-resolution radio, X-ray and/or infrared imaging, as well as integral-field spectroscopy.
For 18 double-peaked AGNs with FIRST detections and optical long-slit spectra that exhibit two relatively compact AGN emission components with angular separations of $\geq$ 0.2 arcsec, \citet{2015ApJ...813..103M} found that $\sim$ 17\% (3/18) galaxies show two spatially separated radio cores in high-resolution multi-band Very Large Array (VLA) images, suggesting these three galaxies as dual AGNs.
\citet{2015ApJ...806..219C} used high-resolution \textit{HST} and \textit{Chandra} observations to investigate 12 galaxies in which the long-slit spectroscopy shows two ${\hbox{[O\,{\sc iii}]}}\lambda$5007 components separated by $\geq$ 0.75 arcsec \citep{2011ApJ...732....9G, 2011ApJ...735...48S,2012ApJ...753...42C}.
They confirmed one dual AGN system where the two stellar bulges have coincident ${\hbox{[O\,{\sc iii}]}}\lambda$5007 and X-ray counterparts, corresponding to $\sim$ 8\% (1/12) of the sample.
Moreover, \citet{2012ApJ...745...67F} investigated 106 double-peaked narrow emission line galaxies using high-resolution Keck/\textit{HST} images or integral-field spectroscopy (IFS) data, suggesting $\sim$ 2\% (2/106) of these galaxies are dual AGNs.
The double-peaked profiles of these two objects are driven by the orbital motion of merging galaxies, with two concentrated ${\hbox{[O\,{\sc iii}]}}\lambda$5007 components spatially coincident with two stellar nuclei.

Considering that the majority of double-peaked AGNs are not expected to be dual AGNs, other mechanisms such as rotating discs and inflows/outflows have been proposed by several studies to explain the origin of double-peaked emission line profiles \citep{2005ApJ...627..721G, 2011ApJ...735...48S, 2012ApJ...752...63S, 2012ApJ...745...67F, 2015ApJ...813..103M, 2016ApJ...832...67N, 2017MNRAS.470.1703D, 2018MNRAS.473.2160N, 2018ApJ...867...66C, 2024MNRAS.534..400B}.
For example, \citet{2015ApJ...813..103M} found that the double-peaked profiles in $\sim$ 72\% (13/18) galaxies are produced by rotating discs or outflows.
They identified radio jet-driven outflows in 5 out of 13 galaxies based on extended radio emission and alignment of the radio jet with the ionized gas.
Additionally, 7 out of 13 galaxies were classified as AGN wind-driven outflows primarily due to significant misalignment ($ > 15^{\circ}$) between the position angle of the two ${\hbox{[O\,{\sc iii}]}}\lambda$5007 emission components and the photometric major axis of the galaxy.
The remaining 1 out of 13 galaxies was classified as a rotating disc, with the two ${\hbox{[O\,{\sc iii}]}}\lambda$5007 emission components spatially coincident with the galaxy's photometric major axis.
In a follow-up study based on the same parent sample of 340 double-peaked AGNs, \citet{2016ApJ...832...67N} used optical long-slit observations to systematically classify 71 double-peaked Type 2 AGNs at z < 0.1 into outflow-dominated and rotation-dominated categories. 
As a result, $\sim$ 86\% (61/71) galaxies with broad wing components were classified as outflow-dominated.
In contrast, $\sim$ 6\% (4/71) galaxies were classified as rotation-dominated. 
In these rotation-dominated galaxies, the ${\hbox{[O\,{\sc iii}]}}\lambda$5007 spectrum shows two narrow components, which are located along the photometric major axis of the galaxy.
The remaining $\sim$ 8\% (6/71) were classified as "Ambiguous", with two narrow components, which are not aligned with the photometric major axis of the galaxy.

However, long-slit spectroscopy has limited spatial coverage, resulting in the loss of spatial information when analyzing spatially extended structures. 
Comparing to the long-slit spectroscopy, integral-field spectroscopy provides more complete spatial coverage and is thus crucial for investigating the nature of double-peaked narrow emission-line galaxies.
With the complete release of the Mapping Nearby Galaxies at Apache Point Observatory \citep[MaNGA;][]{2015ApJ...798....7B} integral field unit (IFU) survey, several works have tried to explore the origin of double-peaked emission line profiles using IFU data, including case studies \citep{2019MNRAS.482.1889W, 2021A&A...653A..47M} and sample-based statistical studies \citep{2023MNRAS.524.5827F, 2024ApJ...976...15Q, 2025arXiv250721512Q}.
From $\sim$ 10,000 galaxies in the final data release of MaNGA, \citet{2023MNRAS.524.5827F} identified 188 galaxies with emission lines that cannot be described by a single Gaussian component.
Among them, 49 galaxies exhibit double-peaked profiles in both the Balmer lines (e.g. H$\alpha$) and forbidden lines (e.g. ${\hbox{[O\,{\sc iii}]}}$$\lambda$5007).
More recently, \citet{2024ApJ...976...15Q} constructed a sample of 304 galaxies with double-peaked H$\alpha$-${\hbox{[N\,{\sc ii}]}}$$\lambda$$\lambda$6548,6583 emission lines in MaNGA, each galaxy containing at least 5 spaxels requiring double Gaussian fitting.
They found that these spaxels with double-peaked emission line structures are statistically associated with bars, AGNs, or tidal features.
However, previous sample-based studies have not separated the two components of double-peaked emission line profiles and have not analyzed the physical properties of each component in detail.

In this work, we select 36 double-peaked narrow emission-line galaxies (DPGs) from the final data release of MaNGA survey and investigate the origin of their double-peaked emission line profiles by performing spatially resolved analyses on the kinematic properties and ionization mechanisms of each component.
Different from \citet{2024ApJ...976...15Q}, we not only require sufficiently strong double-peaked H$\alpha$-${\hbox{[N\,{\sc ii}]}}$$\lambda$$\lambda$6548,6583 emission lines in the spectra, but also require that H$\beta$, ${\hbox{[O\,{\sc iii}]}}$$\lambda$$\lambda$4959,5007, and ${\hbox{[S\,{\sc ii}]}}$$\lambda$$\lambda$6717,6731 be sufficiently strong and exhibit double-peaked profiles, as these emission lines are necessary for BPT diagnostic diagrams.
Additionally, we exclude ongoing mergers and galaxies with a narrow core component plus a broad wing component in the sample selection, which results in a smaller sample size.
The paper is organized as follows.
In Section~\ref{Data analysis}, we give a brief introduction on the MaNGA survey and a detailed description of the sample selection method. 
In Section~\ref{Results}, we analyze the gas kinematics and the ionization mechanisms of the galaxies in our double-peaked sample. 
We discuss the possible origins of the double-peaked emission line profiles in DPGs in Section~\ref{The origin of DPGs}.
Finally, we present our conclusions in Section~\ref{Conclusions}.
We adopt a flat $\Lambda$CDM cosmology with parameters $H_0=70$ $\text{km s}^{-1} \text{ Mpc}^{-1}$, $\Omega_\text{m} = 0.3$, and $\Omega_\Lambda = 0.7$ throughout this paper.

\section{Data analysis}
\label{Data analysis}

\begin{figure*}
\centering
\includegraphics[width=0.9\textwidth]{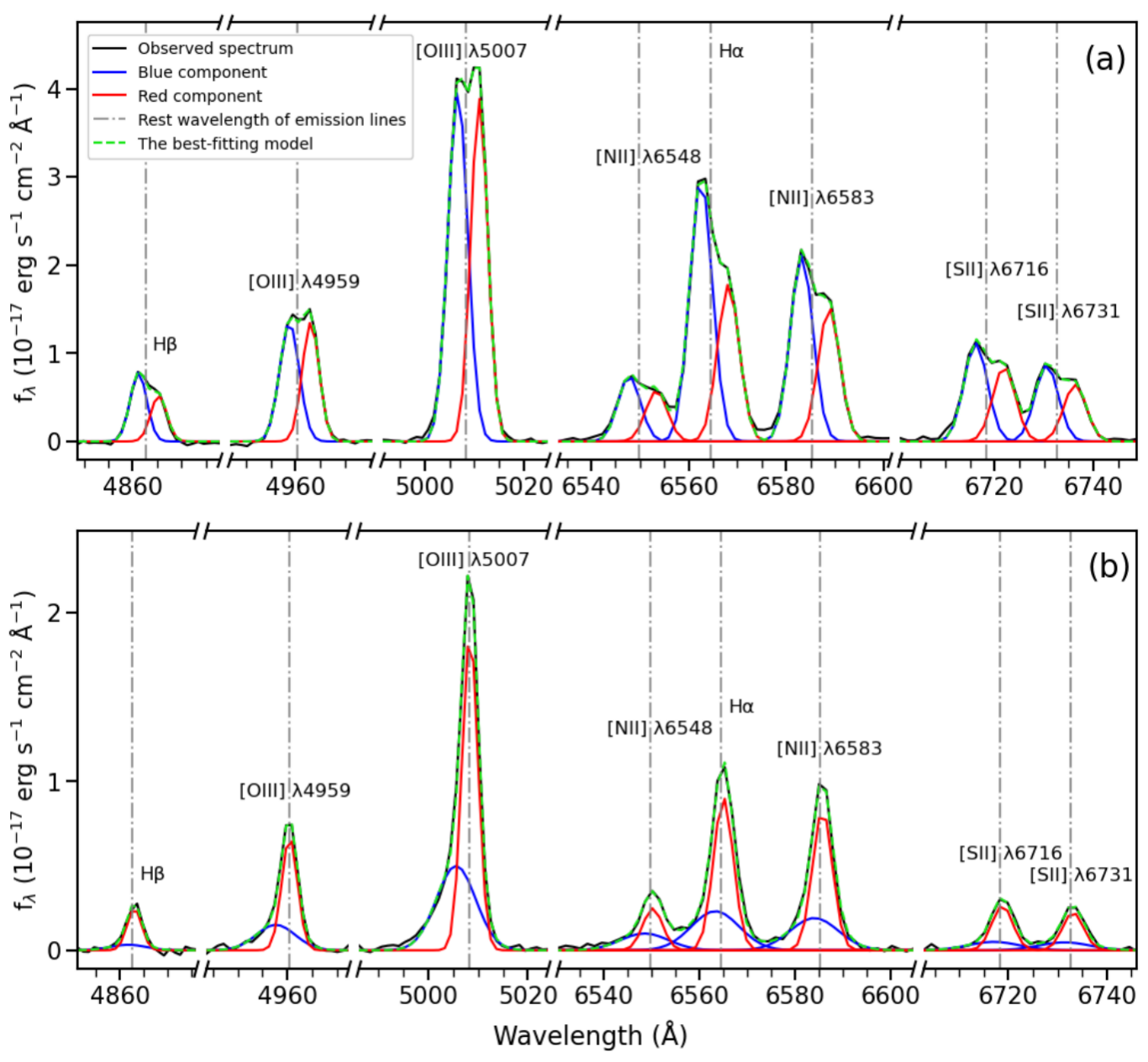}
\centering
\caption{
The emission line spectra from the central spaxel of two example galaxies.
Panel (a): The spectrum of a double-peaked narrow emission-line galaxy (DPG, MaNGA ID: 1-339094). 
Panel (b): The spectrum of a galaxy exhibiting a narrow core and a broad wing in each emission line (MaNGA ID: 1-54940). 
In both panel, The black solid lines show the observed emission line spectra, while the blue and red solid lines represent the blue and red components of the double Gaussian model, respectively.
The green dashed line represents the best-fit model, which is the combination of the blue and red components. 
In addition, the vertical gray dot-dashed lines mark the rest-frame wavelength centre of each emission line.}
\label{Fig.1}
\end{figure*}

\subsection{The MaNGA survey}
MaNGA is one of the three core programs in the fourth-generation Sloan Digital Sky Survey \citep[SDSS-IV;][]{2017AJ....154...28B}, using the 2.5 m Sloan Foundation Telescope at the Apache Point Observatory \citep{2006AJ....131.2332G}.
MaNGA employs dithered observations with 17 hexagonal IFUs \citep{2015AJ....150...19L} with 5 sizes varying between 19 and 127 fibers (or 12.5--32.5 arcsec diameter in the sky) to explore the detailed internal structure of nearby galaxies \citep{2015AJ....149...77D}. 
Two dual-channel BOSS spectrographs \citep{2013AJ....146...32S} provide simultaneous spectral coverage over 3,600--10,300 Å with a median spectral resolution of $R$ $\sim$ 2000 \citep{2016AJ....152...83L}.
The MaNGA project has conducted IFU observations of $\sim$ 10,000 nearby galaxies with a flat stellar mass distribution in 9 $\le$ log ($M_*$/$M_\odot$) $\le$ 11 and a redshift coverage of 0.01 < $z$ < 0.15 \citep{2017AJ....154...86W, 2022ApJS..259...35A}.

The MaNGA Data Reduction Pipeline \citep[DRP;][]{2016AJ....152...83L} provides sky-subtracted and spectrophotometrically-calibrated 3D spectra for each individual galaxy. In addition, the MaNGA Data Analysis Pipeline \citep[DAP;][]{2019AJ....158..231W} analyzes datacube produced by the DRP and provides higher-level data products. 
The MaNGA DAP heavily uses pPXF \citep{2004PASP..116..138C} and a selected subset of stellar templates from the MaSTar library \citep{2019ApJ...883..175Y} to fit the stellar continuum, calculate absorption-line indices, and measure 21 prominent nebular emission lines in the MaNGA wavelength coverage. 
In this work, we use the MaNGA sample and DAP products drawn from SDSS DR17 \citep{2022ApJS..259...35A}, which includes 10,010 unique galaxies.
The MaNGA DAP products named "SPX-MILESHC-MASTARSSP" provide spatially resolved spectral properties, including stellar kinematics, properties of emission lines (e.g. fluxes, velocities, velocity dispersions, and equivalent widths), and spectral indices such as the 4000 Å break ($\text{D}_\textit{n}$4000).
We also measure the global $\text{D}_\textit{n}$4000 for each galaxy from the stacked spectrum. 
We stack the spectra with median signal-to-noise ratio (S/N) per spaxel greater than 2 within the MaNGA bundle.
Additionally, the redshift ($z$), global stellar mass ($M_{*}$), and effective radius ($R_\text{e}$, half-light radius in the $r$-band) are adopted from the NASA Sloan Atlas (NSA) catalog \citep{2011AJ....142...31B}, the Sérsic index $n$ is adopted from MaNGA PyMorph photometric catalogue \citep{2019MNRAS.483.2057F}.

\subsection{Sample selection}
To ensure that the emissions of the ${\hbox{[O\,{\sc iii}]}}$$\lambda$5007 and H$\alpha$ are sufficiently strong, we first select 1,837 out of 10010 galaxies in which the equivalent widths of both the ${\hbox{[O\,{\sc iii}]}}$$\lambda$5007 ($\text{EW}_\text{${\hbox{[O\,{\sc iii}]}}$}$) and H$\alpha$ ($\text{EW}_\text{H$\alpha$}$) in the central spaxel are greater than 3 Å.
We exclude 94 galaxies which show broad H$\alpha$ emission lines in the central spaxel (FWHM $>$ 1000 km s$^{-1}$).

For the remaining 1743 galaxies, we apply both the single and double Gaussian models to fit each emission line for spaxels with both H$\alpha$ and ${\hbox{[O\,{\sc iii}]}}$$\lambda$5007 S/N $>$ 3.

\subsubsection{Single Gaussian model}
\label{Single Gaussian model}
The single Gaussian model is constructed and implemented using the \texttt{curve\_fit} function from the \texttt{scipy.optimize} module in Python \citep{2020NatMe..17..261V}.
In the fitting process, H$\beta$, ${\hbox{[N\,{\sc ii}]}}$$\lambda$$\lambda$6548,6583, and ${\hbox{[S\,{\sc ii}]}}$$\lambda$$\lambda$6717,6731 emission lines are tied to have the same line centre and line width as H$\alpha$ in the velocity space.
We fit the ${\hbox{[O\,{\sc iii}]}}$$\lambda$$\lambda$4959,5007 independently since their profiles do not typically match the profile of the other emission lines \citep{2013ApJ...775..116R}.
Therefore, the ${\hbox{[O\,{\sc iii}]}}$$\lambda$4959 is tied to have the same line centre and line width as ${\hbox{[O\,{\sc iii}]}}$$\lambda$5007 in the velocity space.

The fitting of H$\alpha$, ${\hbox{[N\,{\sc ii}]}}$$\lambda$$\lambda$6548,6583, and ${\hbox{[S\,{\sc ii}]}}\lambda$$\lambda$6717,6731 emission lines is performed in the rest-frame wavelength range 6520Å $\le$ $\lambda$ $ \le$ 6760Å, while the fitting of ${\hbox{[O\,{\sc iii}]}}$$\lambda$$\lambda$4959,5007 is performed in the rest-frame wavelength range 4930Å $\le$ $\lambda$ $ \le$ 5030Å.
Additionally, the H$\beta$ emission line is fitted in the range 4840Å $\le$ $\lambda$ $\le$ 4890Å.
To ensure the fitting is effective for the narrow emission lines, we require the full width at half-maximum (FWHM) of each emission line to be less than 1000 km s$^{-1}$.

\subsubsection{Double Gaussian model}
For the double Gaussian fitting, we closely follow the method described in Section~\ref{Single Gaussian model}, the only difference is that each emission line is modeled by two Gaussian components.

We separate the two Gaussian components of each emission line into blue and red components.
For each emission line, the line centre wavelength of the red component is required to be longer than that of the blue component ($\lambda_\text{blue}$ < $\lambda_\text{red}$).
Fig.~\ref{Fig.1}(a) shows the spectrum from the central spaxel of a DPG.
The black solid line in Fig.~\ref{Fig.1} shows the observed emission line spectrum, while the blue and red solid lines represent the blue and red components of the double Gaussian model, respectively. 
The green dashed line represents the best-fit model, which is the combination of the blue and red components.

\begin{figure}
\centering
\includegraphics[width=0.48\textwidth]{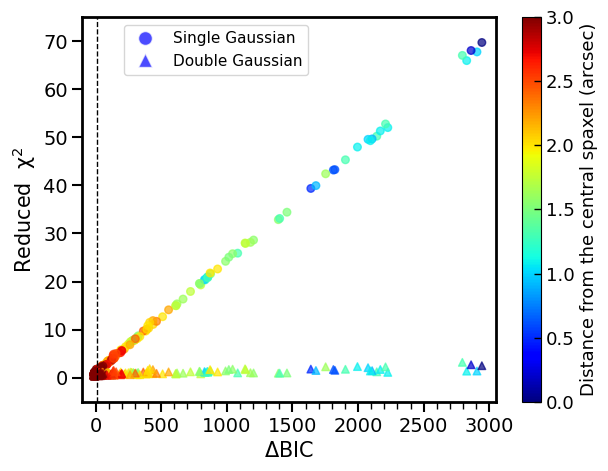}
\centering
\caption{The relation between the $\Delta$BIC and reduced $\chi^{2}$ for single and double Gaussian fittings of ${\hbox{[O\,{\sc iii}]}}$$\lambda$$\lambda$4959,5007 region (4930Å $\le$ $\lambda$ $ \le$ 5030Å) for a DPG example (MaNGA ID: 1-339094).
For the circles, the Y-axis is calculated from the single Gaussian model, while the triangles represent the double Gaussian model.
The data points are colour-coded by the distance to the galaxy centre.
The vertical black dashed line marks the position where $\Delta$BIC = 10, above which the double Gaussian model provides a significantly better fitting than the single Gaussian model.}
\label{Fig.2}
\end{figure}

\subsubsection{Bayesian Information Criterion}
To determine which galaxies require double Gaussian models for emission line fitting, we use the Bayesian Information Criterion \citep[BIC;][]{1978AnSta...6..461S, 2007MNRAS.377L..74L} for both the single ($\text{BIC}_\text{single}$) and double Gaussian models ($\text{BIC}_\text{double}$).
The definition of the BIC is:
\begin{equation}
    \text{BIC = $\chi^{2}$} + k\, \text{ln}\, N,
	\label{eq:BIC}
\end{equation}
where \textit{N} is the number of data points, \textit{k} is the number of free parameters in the model, and $\chi^{2}$ is the chi-square of each model fitting.
For each spaxel, we define $\Delta$BIC as:
\begin{equation}
    \Delta \text{BIC = } \text{BIC}_\text{single} - \text{BIC}_\text{double}.
	\label{deltaBIC}
\end{equation}
According to \citet{2019MNRAS.487..381S} and \citet{2021MNRAS.503.5134A}, $\Delta$BIC $>$ 10 indicates the single Gaussian model is insufficient for accurately describing the emission line profiles, whereas the double Gaussian model provides a significantly better fitting.

Fig.~\ref{Fig.2} shows the relation between the $\Delta$BIC and reduced $\chi^{2}$ for single and double Gaussian fittings of ${\hbox{[O\,{\sc iii}]}}$$\lambda$$\lambda$4959,5007 region for a DPG.
We calculate the reduced $\chi^{2}$ of the single and double Gaussian models respectively.
The circles represent the single Gaussian fittings, and the triangles represent the double Gaussian fittings, colour-coded by the distance to the galaxy centre.
The vertical black dashed line marks the position where $\Delta$BIC = 10, above which the double Gaussian model provides a significantly better fitting than the single Gaussian model.
We find that the difference in reduced $\chi^{2}$ between the two models, as well as $\Delta$BIC, increases with decreasing distance to the galaxy centre.
This suggests that the double Gaussian model is more necessary for the central regions than for the outskirts.

\begin{figure}
\centering
\includegraphics[width=0.48\textwidth]{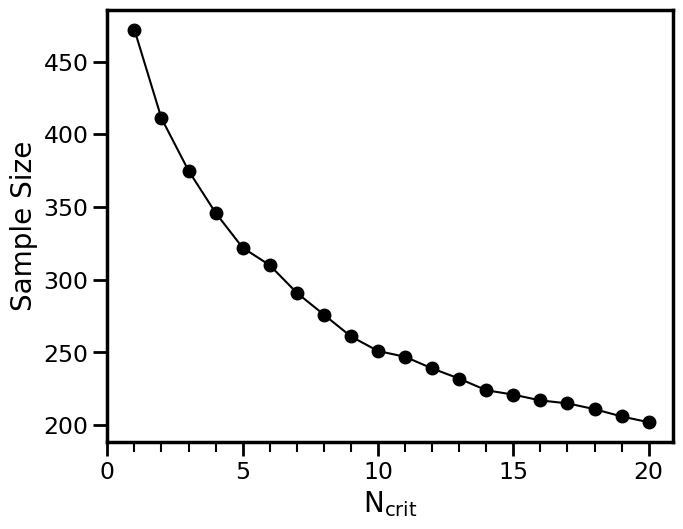}
\centering
\caption{The sample size of double-peaked candidates as a function of the number of contiguous spaxels ($\text{N}_\text{crit}$) that required to satisfying the two criteria of sample selection.}
\label{Fig.3}
\end{figure}

\begin{figure*}
\centering
\includegraphics[width=\textwidth]{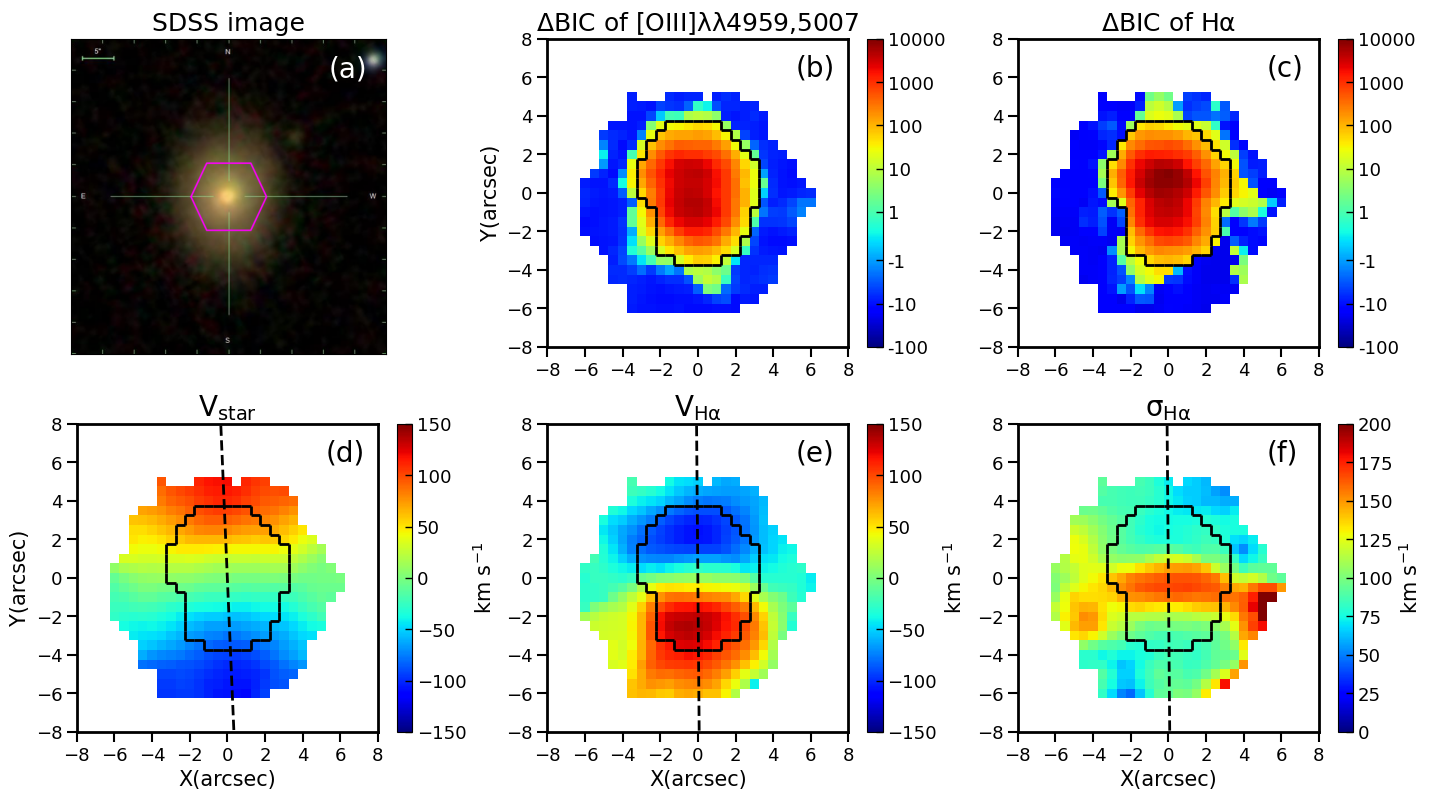}
\caption{
The spatial distributions of $\Delta$BIC,  stellar velocity, H$\alpha$ velocity and H$\alpha$ velocity dispersion of a DPG example (MaNGA ID: 1-339094).
Panel (a): The SDSS \textit{g}-, \textit{r}-, \textit{i}-band image.
Panel (b) \& (c): The $\Delta$BIC maps for ${\hbox{[O\,{\sc iii}]}}$$\lambda$$\lambda$4959,5007 and H$\alpha$ regions, respectively.
Panel (d): The stellar velocity field, the black dashed line represent the kinematic major axis of the stellar component.
Red represents moving away from us while blue represents moving toward us.
Panel (e) \& (f): The velocity field and the velocity dispersion field of H$\alpha$, the black dashed line represent the kinematic major axis of the gaseous component.
Spaxels within the black polygons in panel (b) $\sim$ (f) have $\Delta$BIC $>$ 10 for both the ${\hbox{[O\,{\sc iii}]}}$$\lambda$$\lambda$4959,5007 and H$\alpha$ regions.
}
\label{Fig.4}
\end{figure*}

\subsubsection{Selection of the DPG sample}
For the 1,743 galaxies, we measure $\Delta$BIC for spaxels with both H$\alpha$ and ${\hbox{[O\,{\sc iii}]}}$$\lambda$5007 emission lines S/N $>$ 3.
In order to be classified as a double-peaked candidate, a galaxy is required to have more than 10 contiguous spaxels satisfying the following criteria:

(\textit{i}) $\Delta$BIC $>$ 10 for both the ${\hbox{[O\,{\sc iii}]}}$$\lambda$$\lambda$4959,5007 region and the H$\alpha$ region.
Namely, each emission line should be modelled by a double Gaussian model.

(\textit{ii}) The flux ratio between the blue and red components is in the range of [0.1, 10] for both the ${\hbox{[O\,{\sc iii}]}}$$\lambda$$\lambda$4959,5007 and H$\alpha$ regions
\citep{2009ApJ...705L..76W}.
This flux ratio limit can effectively exclude the influence of the background noise.

Based on these criteria, 225 galaxies are selected as double-peaked candidates.
In fact, the number of double-peaked candidates selected by this method changes with the number of contiguous spaxels ($\text{N}_\text{crit}$) that required to satisfying the above two criteria.
Fig.~\ref{Fig.3} shows the sample size as a function of $\text{N}_\text{crit}$. 

It is clear that the sample size decreases quickly from 472 at $\text{N}_\text{crit}$ $=$ 1 to 225 at $\text{N}_\text{crit}$ $=$ 15. The downward trend slows down at $\text{N}_\text{crit}$ $>$ 15. 
Considering that the typical MaNGA point spread function (PSF) is modeled as a circular Gaussian with FWHM $\sim$ 2.5 arcsec \citep[5 spaxels;][]{2015AJ....150...19L}.
In this work, we use $\text{N}_\text{crit}$ $=$ 15 for the sample selection, corresponding to about three times the MaNGA PSF FWHM, to ensure that the detected features are spatially resolved.

We visually inspect these 225 double-peaked candidates and classify them into two categories:

(\textit{i}) 53 galaxies exhibit double-peaked narrow profiles in ${\hbox{[O\,{\sc iii}]}}$$\lambda$5007, H$\alpha$, as well as other emission lines (H$\beta$, ${\hbox{[O\,{\sc iii}]}}$$\lambda$4959, ${\hbox{[N\,{\sc ii}]}}$$\lambda$$\lambda$6548,6583, and ${\hbox{[S\,{\sc ii}]}}$$\lambda$$\lambda$6717,6731), as shown in Fig.~\ref{Fig.1}(a).
The two components of the double-peaked emission line profiles have similar velocity dispersions ($0.67 < \sigma_\text{blue} / \sigma_\text{red} < 1.5$), where $\sigma_\text{blue}$ and $\sigma_\text{red}$ represent the velocity dispersions of the blue and red components, respectively.

(\textit{ii}) 172 galaxies show a narrow core and a broad wing component, as shown in Fig.~\ref{Fig.1}(b).
The velocity dispersion of the broad wing component is 1.5 times larger than the narrow core component ($\sigma_\text{broad} / \sigma_\text{narrow} > 1.5$).
The broad wing components in these galaxies are suggested as outflow contribution \citep{2005ApJ...627..721G, 2016ApJ...817..108W, 2016ApJ...819..148K}.

In this work, we do not include the outflow candidates in our double-peaked sample. 
For the remaining 53 galaxies exhibiting double-peaked emission line profiles, we visually exclude 17 ongoing mergers, in which the two narrow emission components may be primarily contributed by the gas of the two galaxies in merging, leaving 36 DPGs as our final sample.
Other external effects that could generate double-peaked line profiles are discussed in Section~\ref{External processes in DPGs}.

Fig.~\ref{Fig.4} shows the spatially resolved $\Delta$BIC and the kinematics of a DPG example (MaNGA ID: 1-339094).
Fig.~\ref{Fig.4}(a) shows the SDSS \textit{g}-, \textit{r}-, \textit{i}-band image.
Fig.~\ref{Fig.4}(b) \& (c) show the $\Delta$BIC maps for ${\hbox{[O\,{\sc iii}]}}$$\lambda$$\lambda$4959,5007 and H$\alpha$ regions for spaxels where both the H$\alpha$ and ${\hbox{[O\,{\sc iii}]}}$$\lambda$5007 S/N $>$ 3.
Spaxels within the black polygon in Fig.~\ref{Fig.4} have $\Delta$BIC $>$ 10 for both the ${\hbox{[O\,{\sc iii}]}}$$\lambda$$\lambda$4959,5007 and H$\alpha$ regions.
Fig.~\ref{Fig.4}(d), (e) and (f) show the stellar velocity field, the H$\alpha$ velocity and velocity dispersion fields provided by MaNGA DAP with single Gaussian fitting of each emission line, respectively.
In the velocity fields, red represents moving away from us while blue represents moving toward us.
The black dashed lines represent the kinematic major axes of the stellar (Fig.~\ref{Fig.4}d) and gaseous (Fig.~\ref{Fig.4}e \& f) components fitted by the \texttt{fit\_kinematic\_pa} module in Python \citep{2006MNRAS.366..787K}.

\begin{figure*}
\centering
\includegraphics[width=0.95\textwidth]{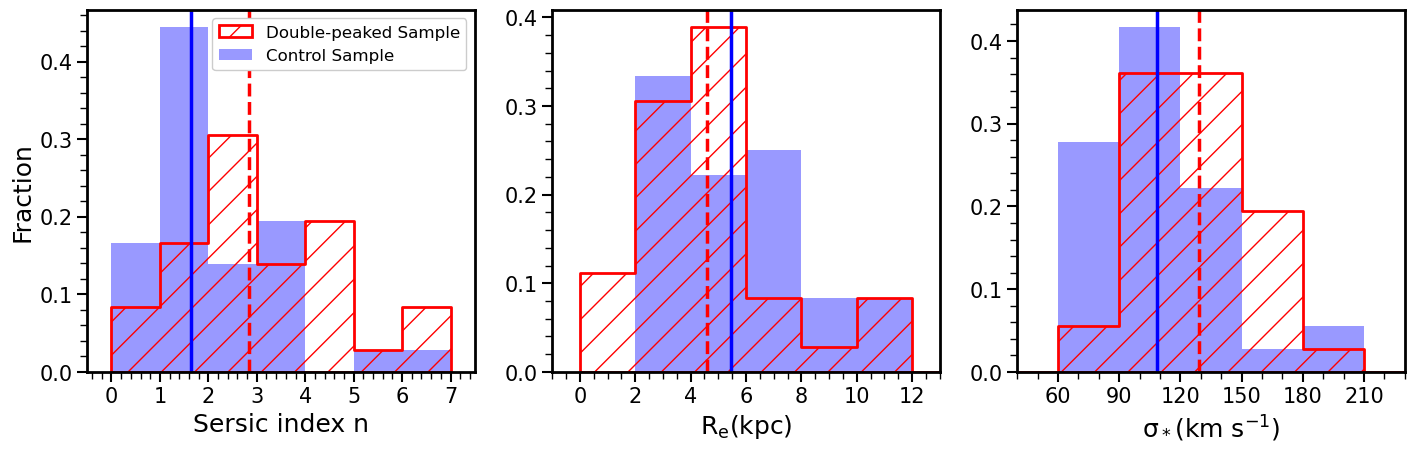}
\caption{Distributions of the Sérsic index ($n$, left), effective radius ($R_\text{e}$, middle) and stellar velocity dispersion ($\sigma_{*}$, right) for the double-peaked sample (red) and the control sample (blue). 
The vertical red dashed and blue solid lines mark the median values of each distribution for the double-peaked sample and the control sample, respectively.
}
\label{Fig.5}
\end{figure*}

\begin{figure*}
\centering
\includegraphics[width=0.98\textwidth]{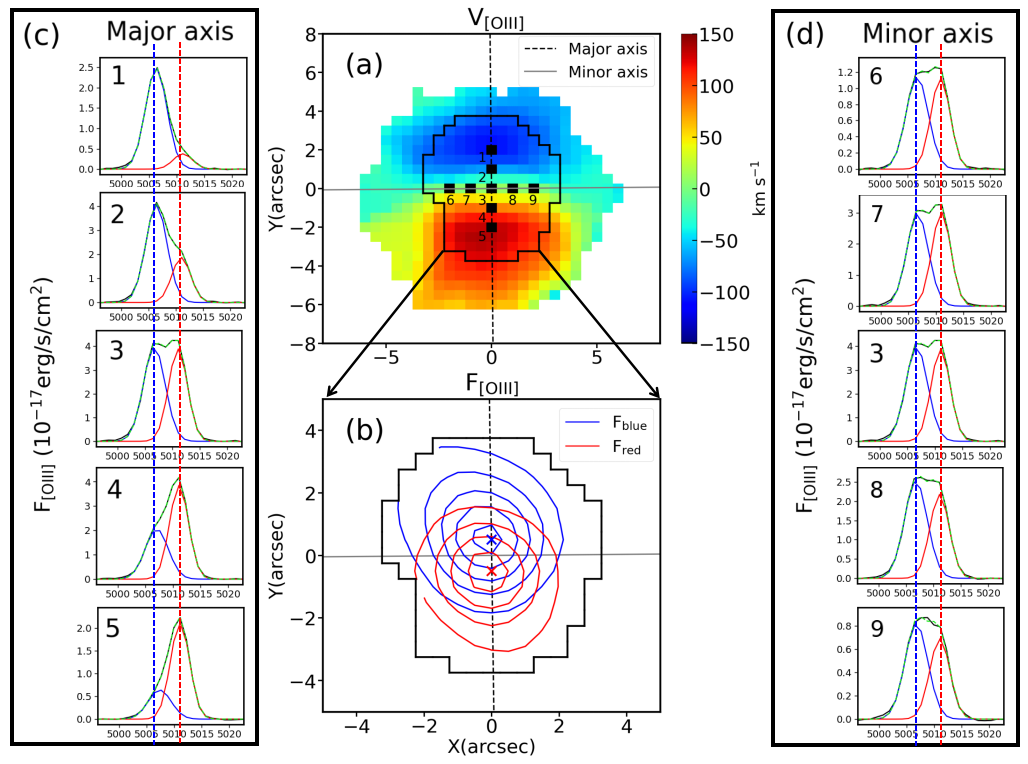}
\caption{
Panel (a): 
The ${\hbox{[O\,{\sc iii}]}}\,\lambda5007$ velocity field of MaNGA 1-339094 provided by the MaNGA DAP based on a single-Gaussian fit.
Spaxels within the black polygon require double Gaussian fitting.
The black dashed line represents the kinematic major axis of the gaseous component, while the grey solid line marks the kinematic minor axis.
The black squares mark the spaxels whose spectra are shown in panel (c) \& (d).
Panel (b): The ${\hbox{[O\,{\sc iii}]}}$$\lambda$5007 flux contours of the blue and red components for spaxels requiring double Gaussian fitting.
The blue and red crosses mark the positions of peak fluxes of the blue and red components, respectively.
Panel (c) \& (d): The ${\hbox{[O\,{\sc iii}]}}$$\lambda$5007 emission line spectra for five spaxels along the major axis (1, 2, 3, 4, 5) and five spaxels along the minor axis (6, 7, 3, 8, 9).
The black solid spectra are the observed emission lines, while the blue and red solid curves represent the blue and red Gaussian components, respectively.
The green dashed line shows the best-fit model, which is the combination of the blue and red components. 
The vertical blue and red dashed lines mark the ${\hbox{[O\,{\sc iii}]}}$$\lambda$5007 line centres of the blue and red components for the central spaxel, spaxel 3.
}
\label{Fig.6}
\end{figure*}

As shown in Fig.~\ref{Fig.4}(d) \& (e), the stellar and gaseous discs are counter-rotating.
As Fig.~\ref{Fig.4}(f) shows, the H$\alpha$ velocity dispersion field exhibits an enhanced region along the minor axis.
In this region, the blue and red components have similar fluxes and show an obvious double-peaked profile. 
The single Gaussian fitting represents the combined contribution of the two components, resulting in a higher velocity dispersion.

\begin{figure*}
\centering
\includegraphics[width=0.98\textwidth]{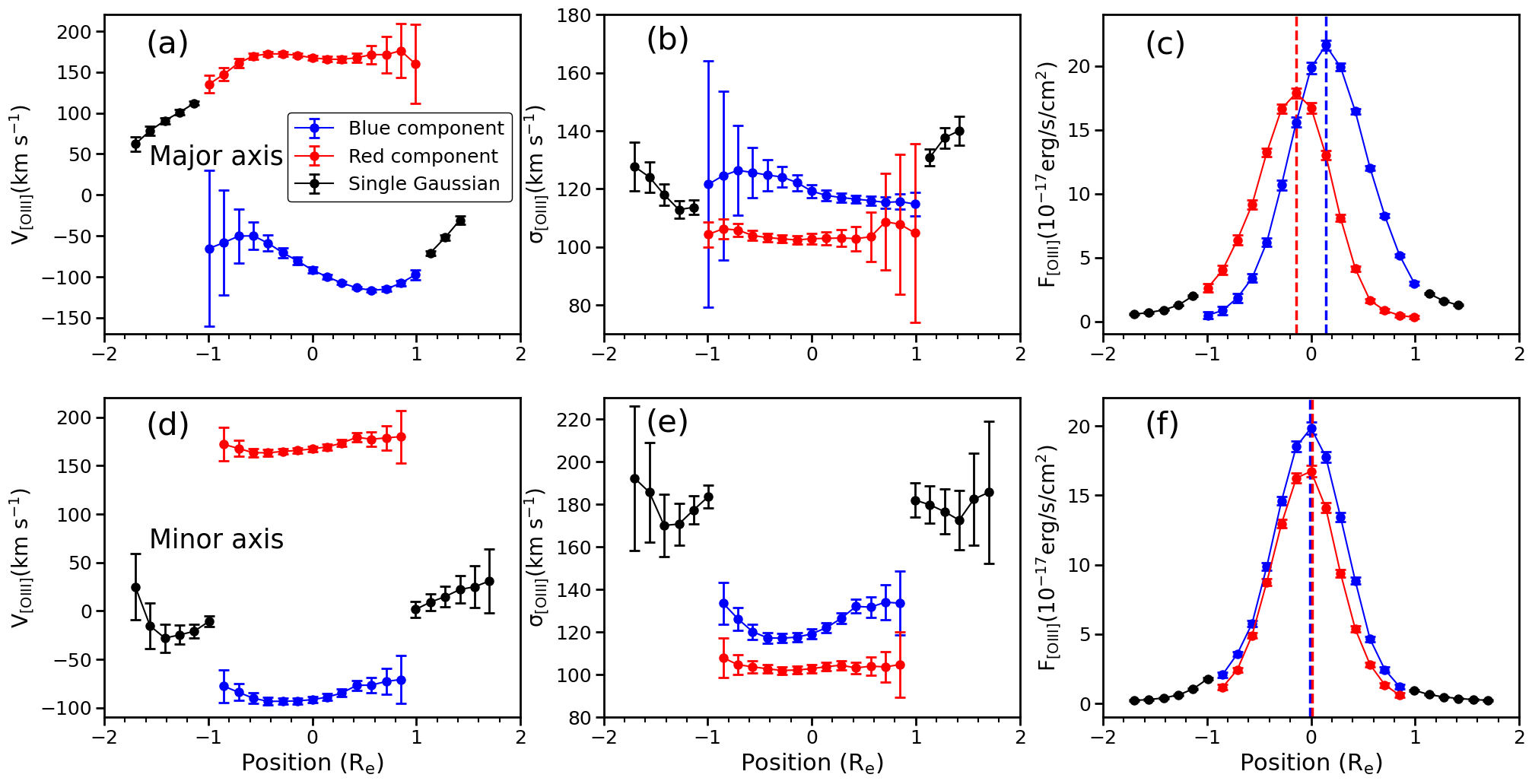}
\caption{
The velocity (left), velocity dispersion (middle), and flux (right) for the blue and red components of the ${\hbox{[O\,{\sc iii}]}}$$\lambda$5007 emission line along the major (top row) and minor (bottom row) axes for MaNGA 1-339094. 
The blue and red dots represent the blue and red components, respectively.
The black dots are the single Gaussian fitting results for spaxels with $\Delta$BIC $<$ 10.
The error bars show the $\pm$ 1$\sigma$ scattering ranges.
The vertical blue and red dashed lines in panel (c) \& (f) mark the positions of the peak fluxes of the blue and red components, respectively.}
\label{Fig.7}
\end{figure*}

\subsection{Control sample}
In order to understand the origin of the double-peak profiles, we build a control sample for comparison.
The control sample is selected from the galaxies with both the $\text{EW}_\text{${\hbox{[O\,{\sc iii}]}}$}$ and $\text{EW}_\text{H$\alpha$}$ $>$ 3 Å for the central spaxel.
For each DPG, we select a control galaxy (CG) whose emission line shows single Gaussian structure.
The criteria for selecting the control sample are summarized below:

(\textit{i}) We stack the spectra of spaxels within a circular aperture of central 1 kpc radius for each galaxy, following the method of \citet{2023A&A...674A..85A}.
The emission line ratios measured from the stacked spectra are required to fall within the same ionization region in the ${\hbox{[S\,{\sc ii}]}}$-BPT diagram \citep{1981PASP...93....5B, 2001ApJ...556..121K} as that of the DPGs based on single Gaussian fitting.
This criterion ensures that the central regions of both the DPGs and their CGs are ionized by the same mechanism.

(\textit{ii}) |$\Delta $log$M_{*}$| $\leq$ 0.1. 
For each DPG, we constrain the corresponding CG to have a similar stellar mass. 
Stellar mass is the most fundamental parameter of a galaxy and is closely associated with many other physical parameters.

(\textit{iii}) |$\Delta \text{D}_\textit{n}$4000| $\leq$ 0.05. 
The CGs and DPGs are constrained to have similar global $\text{D}_\textit{n}$4000 to ensure that they have similar stellar populations.

Fig.~\ref{Fig.5} shows the distributions of the Sersic index ($n$), effective radius ($R_\text{e}$) and the flux-weighted mean stellar velocity dispersion ($\sigma_{*}$) within 1 $R_\text{e}$, for the double-peaked sample (red) and the control sample (blue).
The stellar velocity dispersion is measured by MaNGA DAP.
The vertical red dashed and blue solid lines mark the median values of the each distribution for the double-peaked sample and the control sample.
As shown in the left panel of Fig.~\ref{Fig.5}, the median value of Sérsic index $n$ in our double-peaked sample is $n$ $\sim$ 3, while the median value for the control sample is $n$ $\sim$ 1.5. 
The larger Sérsic index $n$ of double-peaked sample indicating that the DPGs may have larger bulges than CGs.
The middle panel of Fig.~\ref{Fig.5} shows that the median value of $R_\text{e}$ in our double-peaked sample is $\sim$ 1 kpc smaller than the control sample.
The right panel of Fig.~\ref{Fig.5} shows that the median value of $\sigma_{*}$ is $\sim$ 20 km s$^{-1}$ larger than the control sample.

\section{Results}
\label{Results}

\subsection{Gas kinematics}
In this section, we calculate the spatially resolved kinematic parameters and fluxes for the blue and red components of both ${\hbox{[O\,{\sc iii}]}}$$\lambda$5007 and H$\alpha$ through the double Gaussian fitting.
Although we have not tied the line centre and line width in the fitting process, the main results derived from H$\alpha$ are consistent with those from ${\hbox{[O\,{\sc iii}]}}$$\lambda$5007.
Here we only present the results of ${\hbox{[O\,{\sc iii}]}}$$\lambda$5007 fitting as an example.

\begin{figure*}
\centering
\includegraphics[width=0.95\textwidth]{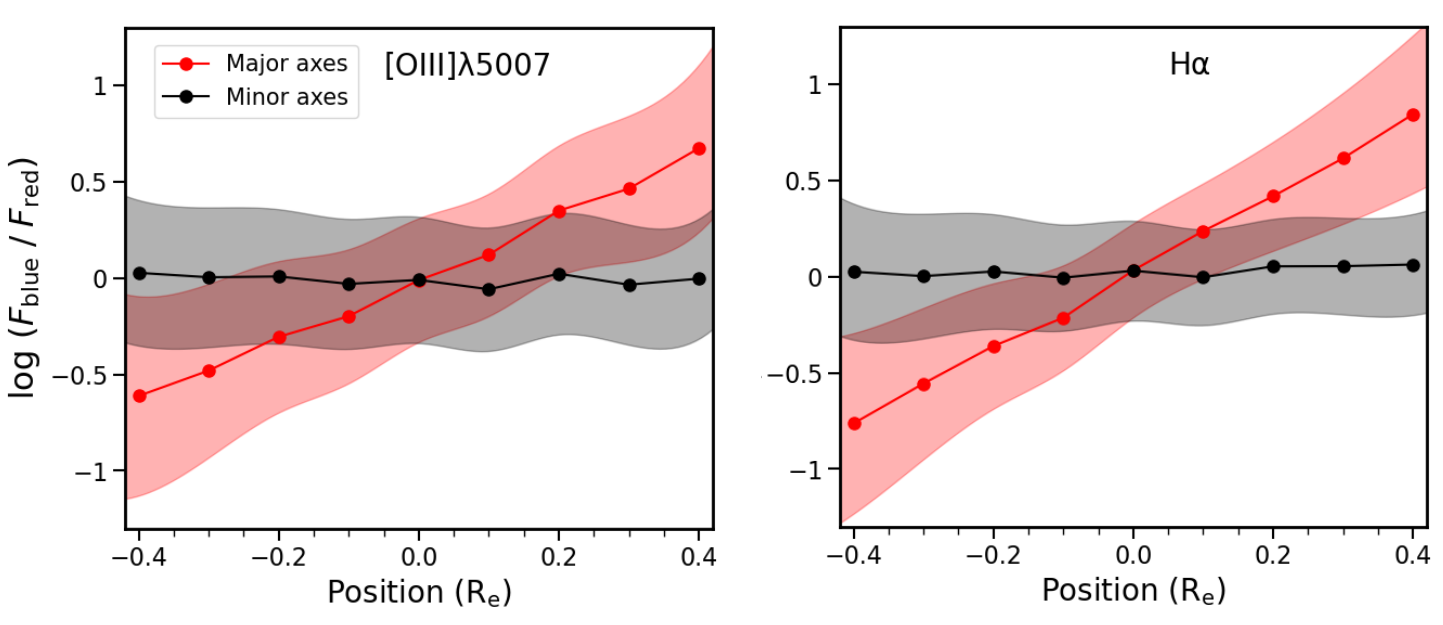}
\caption{
The ${\hbox{[O\,{\sc iii}]}}$$\lambda$5007 (left) and H$\alpha$ (right) flux ratios between the blue and red components along the major and minor axes in DPGs.
For each panel, the red and black dots represent the median of the flux ratios along the major and minor axes for 35 DPGs, respectively.
The position on the X-axis is positive on the blueshifted side of the ${\hbox{[O\,{\sc iii}]}}$$\lambda$5007 velocity field and negative on the redshifted side.
The shadow regions show the $\pm$ 1$\sigma$ scattering ranges.}
\label{Fig.8}
\end{figure*}

\begin{figure*}
\centering
\includegraphics[width=0.98\textwidth]{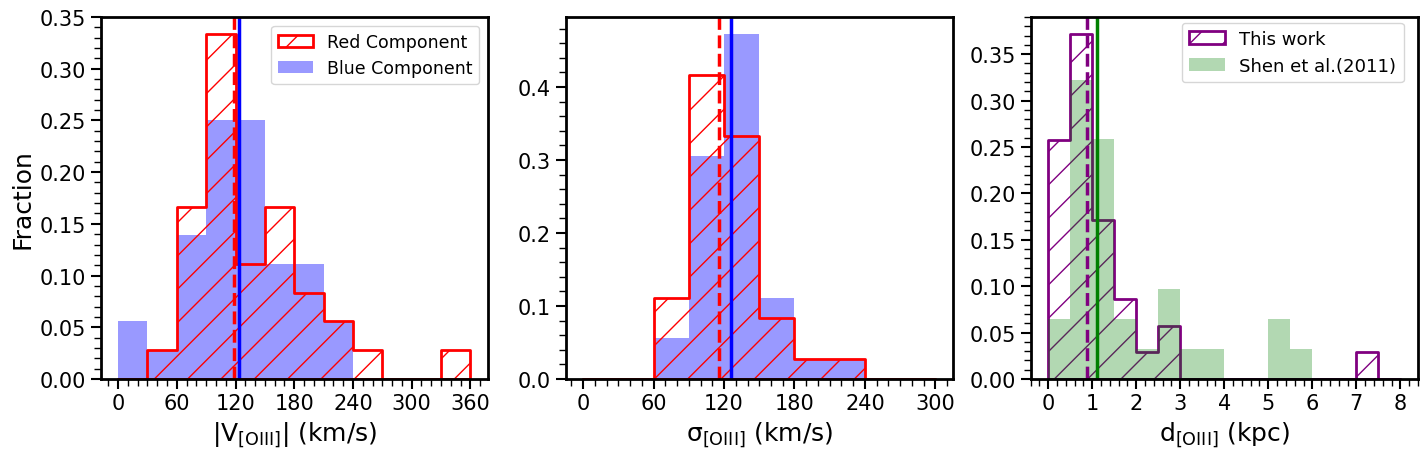}
\caption{The distributions of the line-of-sight velocities (left) and velocity dispersions (middle) for ${\hbox{[O\,{\sc iii}]}}$$\lambda$5007. 
The blue (red) component is shown in blue (red) histogram.
The right panel compares the separation distance of the peak fluxes for the blue and red ${\hbox{[O\,{\sc iii}]}}$$\lambda$5007 components in our work (purple) and that (green) in \citet{2011ApJ...735...48S}. 
The median value of each distribution is marked by the vertical line with the same colour as the distribution.}
\label{Fig.9}
\end{figure*}

Fig.~\ref{Fig.6} shows the result of MaNGA 1-339094 as an example.
Fig.~\ref{Fig.6}(a) shows the ${\hbox{[O\,{\sc iii}]}}$$\lambda$5007 velocity field provided by the MaNGA DAP based on a single-Gaussian fit.
Spaxels within the black polygon require double Gaussian fitting.
The black dashed line represents the kinematic major axis of the gaseous component, while the grey solid line marks the kinematic minor axis.
The black squares mark the spaxels whose spectra are shown in Fig.~\ref{Fig.6}(c) \& (d).
The blue and red contours in Fig.~\ref{Fig.6}(b) show the ${\hbox{[O\,{\sc iii}]}}$$\lambda$5007 flux distributions of the blue ($F_{\text{blue}}$) and red ($F_{\text{red}}$) components for spaxels requiring double Gaussian fitting, respectively.
The blue and red crosses mark the positions of peak fluxes of the blue and red components.
The black dashed and grey solid lines are same as Fig.~\ref{Fig.6}(a).
Fig.~\ref{Fig.6}(c) shows the ${\hbox{[O\,{\sc iii}]}}$$\lambda$5007 emission line spectra for five spaxels along the major axis (1, 2, 3, 4, 5), while Fig.~\ref{Fig.6}(d) shows the spectra for five spaxels along the minor axis (6, 7, 3, 8, 9).
In each spectrum of Fig.~\ref{Fig.6}(c) \& (d), the black solid line represents the observed emission line spectrum, the blue and red solid lines represent the blue and red components, respectively.
The green dashed line shows the best-fit model, which is the combination of the blue and red components. 
The vertical blue and red dashed lines mark the ${\hbox{[O\,{\sc iii}]}}$$\lambda$5007 line centres of the blue and red components for the central spaxel, spaxel 3.
As shown in Fig.~\ref{Fig.6}(c) \& (d), the line centres of both the blue and red components exhibit no significant variations across different spatial positions.

Fig.~\ref{Fig.7} shows the velocity (left), velocity dispersion (middle), and flux (right) for the blue and red components of the ${\hbox{[O\,{\sc iii}]}}$$\lambda$5007 emission line along the major (top row) and minor (bottom row) axes for MaNGA 1-339094.
The blue and red dots represent the blue and red components, respectively.
The black dots are the single Gaussian fitting results for spaxels with $\Delta$BIC $<$ 10.
The error bars show the $\pm$ 1$\sigma$ scattering ranges.
The vertical blue and red dashed lines in Fig.~\ref{Fig.7}(c) \& (f) mark the positions of the peak fluxes of the blue and red components, respectively.

Both Fig.~\ref{Fig.6} and Fig.~\ref{Fig.7}(c) \& (f) show that along the major axis, significant variations can be observed in the line profile and in the flux ratio between the blue and red components; whereas along the minor axis, there is no obvious variation in either the flux ratio or the line profile.

As shown in Fig.~\ref{Fig.6}(c) \& (d) and Fig.~\ref{Fig.7}(a) \& (d), both the blue and red components exhibit no obvious velocity variations along the major or minor axes.
For MaNGA 1-339094, the blue (red) component exhibits a velocity of $\sim$ $-90$ km s$^{-1}$ ($\sim$ $160$ km s$^{-1}$), relative to the stellar velocity of the central spaxel.
In addition, Fig.~\ref{Fig.7}(b) \& (e) show that the velocity dispersions of both the blue and red components keep roughly a constant along either the major or minor axes.
The blue component exhibits a velocity dispersion of $\sim$ $120$ km s$^{-1}$, while the red component has a velocity dispersion of $\sim$ $100$ km s$^{-1}$.

Additionally, Fig.~\ref{Fig.7}(a) \& (c) show that along the major axis, the velocity and flux of the single Gaussian fitting in the outskirts (black dots) are the natural continuation of one of the double Gaussian components for the inner region.
In contrast, Fig.~\ref{Fig.7}(d) \& (f) show that along the minor axis, the velocity and flux of the single Gaussian component in the outskirts are not consistent with either the blue or the red component for the inner region.
Combining Fig.~\ref{Fig.7}(b) \& (e), we find that the velocity dispersion of the single Gaussian fitting along the minor axis is larger than that along the major axis, consistent with the result of enhanced velocity dispersion along the minor axis in Fig.~\ref{Fig.4}(f).

From Fig.~\ref{Fig.6}(b), we find that the flux distributions of the blue and red components deviate from each other along the major axis.
The vertical blue and red dashed lines in Fig.~\ref{Fig.7}(c) \& (f) show that along the major axis, the peak fluxes of the blue and red components have a separation of $\sim$ 0.3 $R_\text{e}$, which corresponds to $\sim$ 0.65 kpc; whereas along the minor axis, the positions of the peak fluxes of the two components overlap with each other.
In addition, Fig.~\ref{Fig.7}(f) shows that the two components have comparable fluxes along the minor axis, leading to the enhanced velocity dispersion observed in the single Gaussian fitting shown in Fig.~\ref{Fig.7}(e).

\begin{figure}
\centering
\includegraphics[width=0.49\textwidth]{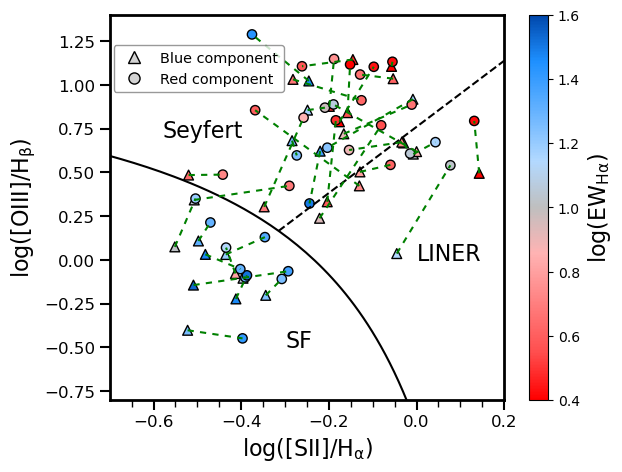}
\caption{
The ${\hbox{[S\,{\sc ii}]}}$-BPT diagram for 
the blue (triangles) and red (circles) components measured from the stacked spectra of the central 1 kpc circular region.
The data points are colour-coded by the equivalent width of H$\alpha$ ($\text{EW}_\text{H$\alpha$}$).
The black solid curve \citep{2001ApJ...556..121K} represents the demarcation line that separate AGN and star-forming (SF) galaxies/regions. 
The black dashed straight line \citep{2006MNRAS.372..961K} distinguishes Seyferts from LINERs.
In addition, the dashed green lines connect the two components of the same galaxy.}
\label{Fig.10}
\end{figure}

The analysis above focuses on the individual DPG.
To further explore whether similar kinematic properties are statistically present in DPGs, we measure the ${\hbox{[O\,{\sc iii}]}}$$\lambda$5007 and H$\alpha$ flux ratios between the blue and red components along the major and minor axes for all 36 DPGs.
We find that 35 out of these 36 DPGs show similar properties as MaNGA 1-339094.
Fig.~\ref{Fig.8} shows the ${\hbox{[O\,{\sc iii}]}}$$\lambda$5007 (left) and H$\alpha$ (right) flux ratios between the two components along the major and minor axes for these 35 DPGs.
In each panel, the red and black dots represent the median flux ratios along the major and minor axes, respectively
The position on the X-axis is positive on the blueshifted side of the ${\hbox{[O\,{\sc iii}]}}$$\lambda$5007 velocity field and negative on the redshifted side.
The shadow regions show the $\pm$ 1$\sigma$ scattering ranges.
Both the median flux ratios between the two components of ${\hbox{[O\,{\sc iii}]}}$$\lambda$5007 and H$\alpha$ show a steep gradient along the major axes.
In contrast, there are no obvious variations in the median flux ratios along the minor axes.

Fig.~\ref{Fig.9} shows the distributions of line-of-sight velocities (left) and velocity dispersions (middle) of the blue and red ${\hbox{[O\,{\sc iii}]}}$$\lambda$5007 components. 
The median value of each distribution is marked with a vertical line of the same colour.
It is clear that the median values of the line-of-sight velocities and the velocity dispersions for the red and blue components are similar.
This similar kinematic properties of the two Gaussian components indicate that the two emission line clouds have the same origin.
The right panel compares the separation distance ($\text{d}_\text{${\hbox{[O\,{\sc iii}]}}$}$) of the peak fluxes for the blue and red ${\hbox{[O\,{\sc iii}]}}$$\lambda$5007 components in our work (purple) and that (green) in \citet{2011ApJ...735...48S}.
The vertical purple dashed and green solid lines mark the median values of each distribution.
Most DPGs in our sample have a spatial separation $<$ 3 kpc, with a median value of $\sim$ 1 kpc, consistent with that given in \citet{2011ApJ...735...48S}.

\subsection{Ionization mechanisms}
Mapping diagnostic line ratios across a galaxy can give us insights into the ionization state of the gas.
The widely used system for spectral classification of emission-line galaxies is based on the diagnostic diagrams originally suggested by \citet[][generally referred to as BPT diagrams]{1981PASP...93....5B}.
Standard BPT diagnostic diagrams rely on the emission line ratios ${\hbox{[O\,{\sc iii}]}}$$\lambda$5007/H$\beta$ versus ${\hbox{[O\,{\sc i}]}}$$\lambda$6300/H$\alpha$, ${\hbox{[S\,{\sc ii}]}}$$\lambda$$\lambda$6717,6731/H$\alpha$ or ${\hbox{[N\,{\sc ii}]}}$$\lambda$6583/H$\alpha$ \citep{2001ApJ...556..121K}.
In both ${\hbox{[O\,{\sc i}]}}$$\lambda$6300 and ${\hbox{[S\,{\sc ii}]}}$$\lambda$$\lambda$6717,6731 BPT diagrams, high ionization Seyfert regions/galaxies and low-ionization narrow emission-line regions 
\citep[LINERs;][]{2008MNRAS.391L..29S}
have been shown to be located on two distinct sequences. In the current work, we choose to use the ${\hbox{[S\,{\sc ii}]}}$-BPT diagram since the S/N of ${\hbox{[S\,{\sc ii}]}}$$\lambda$$\lambda$6717,6731 is much higher than ${\hbox{[O\,{\sc i}]}}$$\lambda$6300 in most spectra.

Fig.~\ref{Fig.10} shows the ${\hbox{[S\,{\sc ii}]}}$-BPT diagrams for the blue (triangles) and red (circles) components measured from the stacked spectra of the central 1 kpc circular region.
The data points are colour-coded by the equivalent width of H$\alpha$ \citep[$\text{EW}_\text{H$\alpha$}$;][]{2020ARA&A..58...99S, 2023MNRAS.524.5640R}.
The dashed green lines connect the two components of the same galaxy.
The black solid curve \citep{2001ApJ...556..121K} is the demarcation line that separates the AGN and star-forming (SF) galaxies/regions, and the black dashed straight line \citep{2006MNRAS.372..961K} distinguishes Seyferts from LINERs.

As shown in Fig.~\ref{Fig.10}, SF DPGs exhibit significantly larger $\text{EW}_\text{H$\alpha$}$ than Seyfert/LINER DPGs.
For 83.3\% (30/36) DPGs, both the blue and red components are located in the same ionization region in the ${\hbox{[S\,{\sc ii}]}}$-BPT diagram and exhibit similar $\text{EW}_\text{H$\alpha$}$, which suggests that the two components are ionized by similar sources.
Specifically, 41.7\% (15/36) of the DPGs have both blue and red components classified as Seyfert, 13.9\% (5/36) as LINER, and 27.7\% (10/36) as SF. 
The remaining 16.7\% (6/36) DPGs have their blue and red components located in different ionization regions.
The main results keep the same if we use ${\hbox{[N\,{\sc ii}]}}$-BPT instead of the ${\hbox{[S\,{\sc ii}]}}$-BPT.
The high fraction of AGN in the DPG sample could be due to a relatively high proportion of tidal interaction \citep{2024MNRAS.527.6722R,2023NatAs...7..463R} or because of a selection effect that we require high $\text{EW}_\text{${\hbox{[O\,{\sc iii}]}}$}$ for sample selection.

\section{The origin of DPGs}
\label{The origin of DPGs}
In this section, we discuss the possible origins of the double-peaked emission line profiles of our sample based on observational results listed in Section~\ref{Results}.

\subsection{Dual AGNs}
According to the definition of dual AGNs, both the blue and red components are required to be located in the Seyfert or LINER region of the BPT diagram \citep{2009ApJ...705L..76W}.
In the ${\hbox{[S\,{\sc ii}]}}$-BPT diagram, 41.7\% of DPGs have both the blue and red components classified as Seyfert, and 13.9\% are classified as LINER.
LINERs were first proposed by \citet{1980A&A....87..152H}, and initially classified as low-luminosity AGNs.
Recently, several studies based on IFU observations suggest that the physical size of the LINER regions can extend to kpc scales, and most galaxies with LINER regions are unlikely to be low-luminosity AGNs \citep{2010MNRAS.402.2187S, 2016MNRAS.461.3111B}.
This indicates that the origin of double-peaked emission line profiles is not necessarily associated with AGN activities.

Additionally, \citet{2013MNRAS.429.2594B} suggested that with insufficient spatial resolution in photometric observations, a dual AGN system separated by $\sim$ kpc cannot be resolved, resulting in a measured photometric major axis of the host galaxy aligning with the orientation connecting the two AGNs.
\citet{2016ApJ...832...67N} simulated the long-slit observational features of a dual AGN model. 
They find that along the orientation connecting the two AGNs, the flux ratio between the two narrow emission line components shows systematic variations; while perpendicular to this orientation, the flux ratio shows almost no variation.

Although we find that the flux ratio between the blue and red components varies systematically along the major axes and keeps roughly a constant along the minor axes in 97.2\% (35/36) DPGs (see Fig.~\ref{Fig.8}), we do not suggest the double-peaked line profiles are dominated by dual AGNs.
As known from the literatures, only a very small fraction \citep[$\sim$ 2\%;][]{2012ApJ...745...67F} of double-peaked AGNs are expected to be dual AGNs.

\subsection{Rotating discs}
\label{Rotating discs}
If the double-peaked emission line profiles originate from a rotating disc, we would expect the flux ratio between the two components varies along the major axis, while this ratio keeps a constant along the minor axis \citep{2011ApJ...735...48S, 2015ApJ...813..103M, 2016ApJ...832...67N, 2023A&A...670A..46M}.
In the central regions, the two components have comparable fluxes.
Towards the outskirts along half major axis, one component gradually dominates the line profile, while the other weakens until it becomes undetectable.
This expectation is totally consistent with the observational results shown in Fig.~\ref{Fig.6} and Fig.~\ref{Fig.8}.
Additionally, Fig.~\ref{Fig.7}(a) \& (c) show that along the major axis, the velocity and flux of the single Gaussian component in the outskirts are the natural continuation of the blue or red ${\hbox{[O\,{\sc iii}]}}$$\lambda$5007 component for the inner region.

Furthermore, the two components of the double-peaked profile resulting from rotating discs show similar value of line-of-sight velocities and similar velocity dispersions according to the axisymmetric models in \citet{2023A&A...670A..46M}.
As shown in Fig.~\ref{Fig.9}, the blue and red components of our sample exhibit similar distributions in both the value of line-of-sight velocity and velocity dispersion. 
Thus, we suggest that the double-peaked emission line profiles
in 35 out of 36 DPGs primarily originate from rotating discs.

In order to examine whether dynamic disturbances present in the rotating discs, we check the flux ratio between the blue ($F_{\text{blue}}$) and red ($F_{\text{red}}$) components of both H$\alpha$ and ${\hbox{[O\,{\sc iii}]}}$ emission lines for the central spaxel of each galaxy.
22.9\% (8/35) DPGs have 0.75 $\leq$ $F_{\text{blue}}/F_{\text{red}}$ $\leq$ 1.25 for both H$\alpha$ and ${\hbox{[O\,{\sc iii}]}}$ emission lines. 
We suggest undisturbed rotating discs as the origin of double-peaked emission line profiles in these galaxies.
In contrast, 77.1\% (27/35) DPGs have $F_{\text{blue}}/F_{\text{red}}$ $>$ 1.25 or $F_{\text{blue}}/F_{\text{red}}$ $<$ 0.75 in the central spaxel, these asymmetric line profiles might be due to dynamic disturbances in the rotating discs.

\subsection{Outflows}
\label{outflows}
Galactic-scale outflows are known to be powered by star formation or AGN activities.
In actively star-forming galaxies, galactic-scale outflows are driven by the mechanical energy and momentum from supernovae and stellar winds \citep{1990ApJS...74..833H}.
Young star clusters create overpressured bubbles of hot gas that expand and sweep up surrounding interstellar medium (ISM) until they “blow out” from the disc into the halo.
The collective action of multiple superbubbles results in a weakly collimated biconical outflow along the minor axis of the galaxy \citep{2010AJ....140..445C, 2014ApJ...797...90S}.
In AGN host galaxies, galactic-scale outflows are driven by wide-angle winds launched from the accretion disc, accelerated by thermal energy, radiation pressure, or radio jets \citep{2012ARA&A..50..455F, 2019MNRAS.490.3025R, 2020A&ARv..28....2V}.

\begin{figure}
\centering
\includegraphics[width=0.49\textwidth]{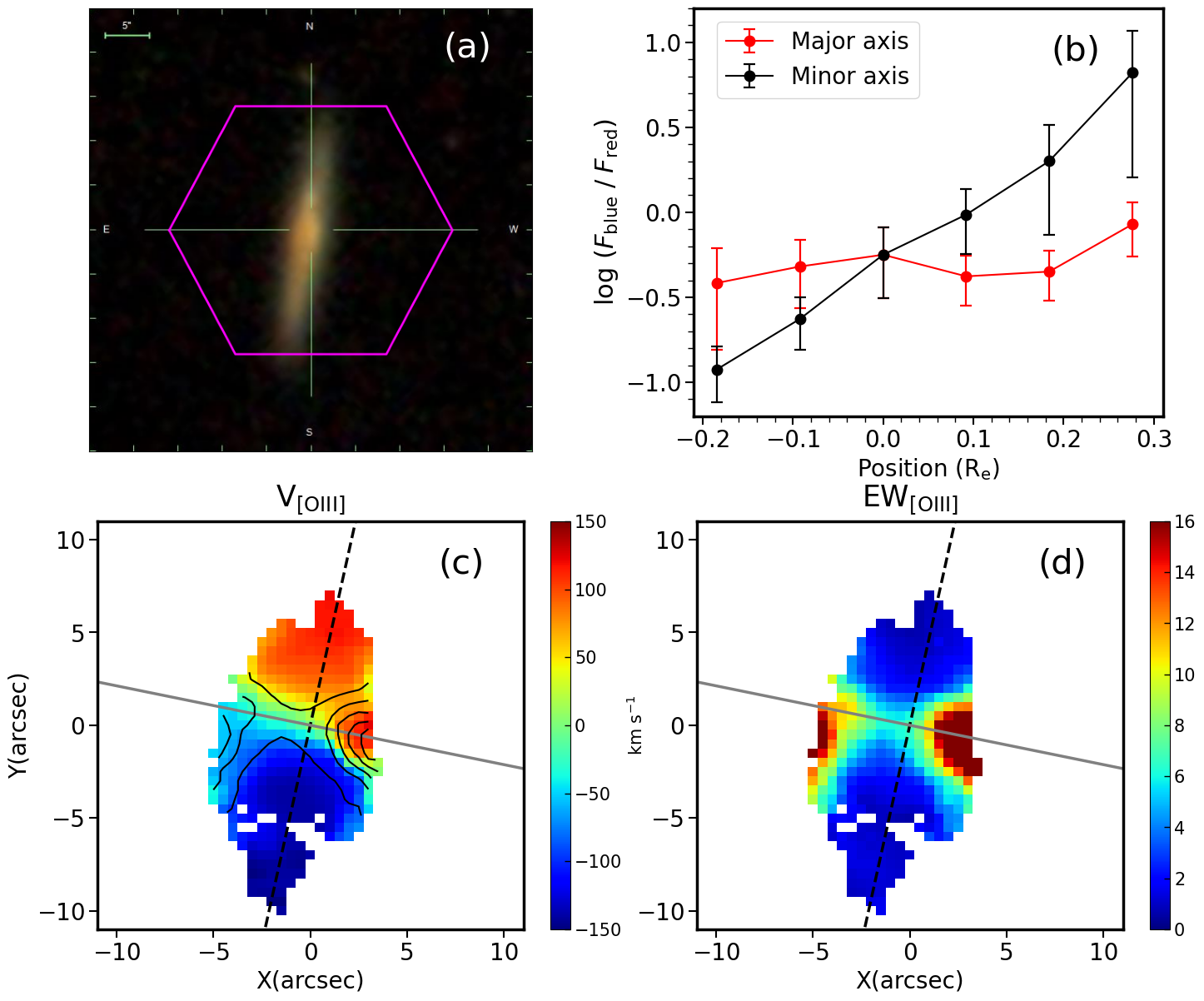}
\caption{Panel (a): The SDSS \textit{g}-, \textit{r}-, \textit{i}-band image of the DPG with biconical outflow (MaNGA ID: 1-210020).
Panel (b): The ${\hbox{[O\,{\sc iii}]}}$$\lambda$5007
flux ratios between the blue and red components along the major (red) and minor axes (black).
The error bars show the $\pm$ 1$\sigma$ scattering ranges.
The position on the X-axis is positive on the blueshifted side of the ${\hbox{[O\,{\sc iii}]}}$$\lambda$5007 velocity field and negative on the redshifted
side.
Panel (c) \& (d): The velocity map and the equivalent width map of ${\hbox{[O\,{\sc iii}]}}$$\lambda$5007, the black dashed line represent the kinematic major axis of the gaseous component, while the grey solid line mark the minor axis.
The black contours in panel (c) mark the regions with enhanced $\text{EW}_\text{${\hbox{[O\,{\sc iii}]}}$}$.}
\label{Fig.11}
\end{figure}

Our results of Fig.~\ref{Fig.6} and Fig.~\ref{Fig.8} are inconsistent with the biconical outflow model, in which the flux ratio between the blue and red components is expected to vary significantly along the minor axis of the galaxy \citep{2017ApJ...834...30F}.
However, we identify one DPG (MaNGA ID: 1-210020) whose ${\hbox{[O\,{\sc iii}]}}$$\lambda$5007 flux ratio between the two Gaussian components shows significant variation along the minor axis, while it keeps roughly a constant along the major axis as shown in Fig.~\ref{Fig.11}.
Fig.~\ref{Fig.11} shows that this DPG is an edge-on galaxy with a biconical ionized structure in the $\text{EW}_\text{${\hbox{[O\,{\sc iii}]}}$}$ map along the minor axis.
Fig.~\ref{Fig.11}(a) shows the SDSS \textit{g}-, \textit{r}-, \textit{i}-band image.
Fig.~\ref{Fig.11}(b) shows the ${\hbox{[O\,{\sc iii}]}}$$\lambda$5007 flux ratios between the blue and red components along the major (red) and minor axes (black).
The error bars show the $\pm$ 1$\sigma$ scattering ranges.
The position on the X-axis is positive on the blueshifted side of the ${\hbox{[O\,{\sc iii}]}}$$\lambda$5007 velocity field and negative on the redshifted side.
Fig.~\ref{Fig.11}(c) \& (d) show the velocity map and the equivalent width map of ${\hbox{[O\,{\sc iii}]}}$$\lambda$5007, the black dashed line represent the kinematic major axis of the gaseous component, while the grey solid line mark the minor axis.
The black contours in panel (c) mark the regions with enhanced $\text{EW}_\text{${\hbox{[O\,{\sc iii}]}}$}$.
The velocity field within the two $\text{EW}_\text{${\hbox{[O\,{\sc iii}]}}$}$ enhanced regions clearly deviates from regular disk rotation due to outflow contribution \citep{2019MNRAS.490.3830B, 2021MNRAS.505..191B, 2024MNRAS.531.2462Z}.
Thus, we suggest that the origin of double-peaked profiles in this DPG is associated with biconical outflow.

\subsection{External processes in DPGs}
\label{External processes in DPGs}
In 77.1\% of DPGs in our sample, the emission lines show asymmetric profiles in the central spaxel due to flux differences between the blue and red components. 
These asymmetric profiles suggest the possibility of dynamic disturbances within these galaxies \citep{2016ApJ...832...67N}.
In this section, we try to figure out the origin of the disturbances.

\begin{figure}
\centering
\includegraphics[width=0.45\textwidth]{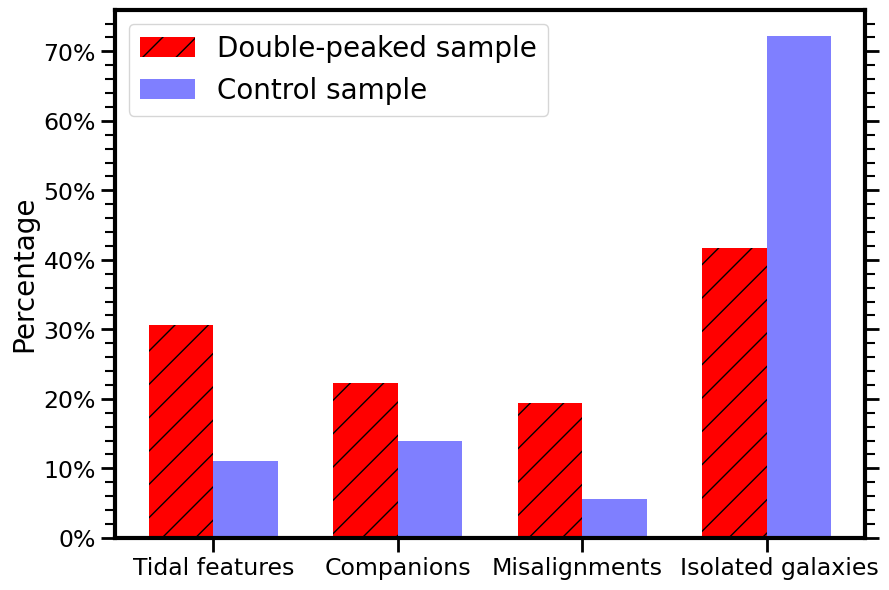}
\caption{Classification of galaxies 
with different features for double-peaked sample (red histograms) and their control sample (blue histograms).}
\label{Fig.12}
\end{figure}

\begin{figure*}
\centering
\includegraphics[width=0.88\textwidth]{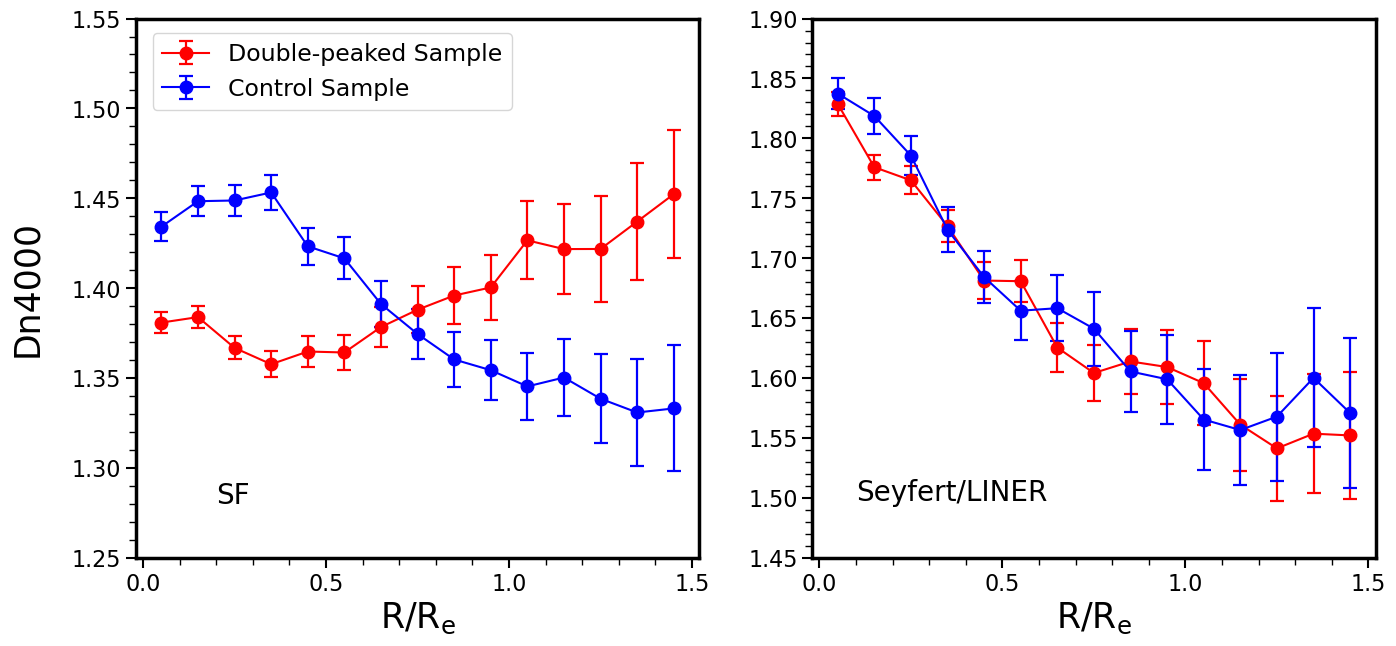}
\caption{
$\text{D}_\textit{n}$4000 radial gradients for SF (left) and Seyfert/LINER (right) galaxies.
The red and blue dots represent the median $\text{D}_\textit{n}$4000 for the double-peaked sample and control sample, respectively.
The error bars are the typical measurement error of $\text{D}_\textit{n}$4000 at different radii.}
\label{Fig.13}
\end{figure*}

We search for evidence of external processes (including tidal features, close companion galaxies, as well as gas-star misalignments) that happened in DPGs as well as their controls.
To search the tidal features in DPGs and their control sample, we visually inspect the images from Dark Energy Spectroscopic Instrument (DESI) Legacy Imaging Surveys \citep{2019AJ....157..168D}, which is $\sim$ 2 mag deeper than SDSS.
We identify galaxies exhibiting tidal tails, shells, as well as faint stellar streams, and consider these galaxies to have tidal features.
We obtain the spectroscopic redshifts of DPGs/controls and their close companions from SDSS single-fiber spectra.
The close companions are defined as galaxies within a projected separation of $r_\textit{p}$ $\leq$ 80 $h_{70}^{-1}$ kpc and a line-of-sight velocity difference $\Delta$V $\leq$ 300 km s$^{-1}$ relative to the DPGs/controls \citep{2013MNRAS.430.3128E}.
We also check the stellar and H$\alpha$ velocity field provided by MaNGA DAP for each galaxy to identify gas-star misaligned galaxies \citep[$|\text{PA}_\text{star}-\text{PA}_\text{gas}| > 30^\circ$; ][]{2022MNRAS.515.5081Z}.
Fig.~\ref{Fig.12} compares the classifications between double-peaked sample (red histograms) and their control sample (blue histograms).
We find that 58.3\% (21/36) of the galaxies in our double-peaked sample exhibit evidence of external processes (including tidal features, close companion galaxies, as well as gas-star misalignments), whereas only 27.8\% (10/36) of the galaxies in the control sample present these characteristics. 
This suggests that the origin of double-peaked emission line profiles is associated with external processes.

We compare the stellar population gradients between the DPGs and their controls using continuum spectral indices 4000 Å break ($\text{D}_\textit{n}$4000).
Fig.~\ref{Fig.13} shows the $\text{D}_\textit{n}$4000 radial gradients for SF (left) and Seyfert/LINER (right) galaxies.
The red and blue dots represent the median $\text{D}_\textit{n}$4000 for the double-peaked sample and control sample, respectively. 
The error bars are the typical measurement error of $\text{D}_\textit{n}$4000 at different radii.
For the SF galaxies, we find that the $\text{D}_\textit{n}$4000 gradient of the double-peaked sample is positive, indicating that the double-peaked sample has younger stellar population in the central regions than their outskirts.
The $\text{D}_\textit{n}$4000 gradient of the control sample is negative, indicating older stellar populations in the central regions, as expected for ordinary bulge + disc structure galaxies.
For the Seyfert/LINER galaxies, the double-peaked sample and the control sample have similarly negative $\text{D}_\textit{n}$4000 gradients.
The positive $\text{D}_\textit{n}$4000 radial gradient for double-peaked SF galaxies can be naturally explained by external processes which trigger gas inflows, leading to a fast centrally concentrated star formation, while shutting down star formation in the outskirts due to the lack of cold gas \citep{2022MNRAS.511.4685X}.
Furthermore, the larger Sérsic index $n$, the higher stellar velocity dispersion, and the smaller $R_\text{e}$, compared to the control sample shown in Fig.~\ref{Fig.5} can also be explained by the gas inflow picture \citep{2020A&A...641A.171M, 2022MNRAS.515.5081Z}.

\section{Conclusions}
\label{Conclusions}
In this work, we study 36 double-peaked narrow emission-line galaxies selected from 10,010 unique galaxies in MaNGA survey. 
We use a double Gaussian model to separate the double-peaked profiles for each emission line and perform spatially resolved analyses on the kinematic properties and ionization mechanisms of each Gaussian component.
The main findings are summarized below:

(\textit{i}) Comparing to the control sample, DPGs exhibit a larger Sérsic index $n$, higher stellar velocity dispersion, and smaller $R_\text{e}$.

(\textit{ii}) In 97.2\% (35/36) DPGs, the blue and red components exhibit two distinct velocities and are spatially offset by $\sim$ kpc scale along the major axes, resulting in significant variations in the flux ratio between the blue and red components along the major axes, while the flux ratio keeps roughly a constant along the minor axes.

(\textit{iii}) The blue and red components of DPGs exhibit similar distributions in both the value of line-of-sight velocity and the velocity dispersion, indicating that the two components have similar kinematic properties and physical origins.

(\textit{iv}) 83.3\% (30/36) DPGs have both blue and red components located in the same ionization region in the ${\hbox{[S\,{\sc ii}]}}$-BPT diagram, suggesting that the two components are ionized by similar sources.
Specifically, 41.7\% DPGs have both components classified as Seyfert, while 13.9\% are classified as LINER and 27.7\% are classified as SF.
The remaining 16.7\% DPGs have their blue and red components located in different ionization regions.

(\textit{v}) For the double-peaked SF galaxies, the $\text{D}_\textit{n}$4000 radial gradient is positive, whereas it is negative for the control sample, indicating that the double-peaked SF galaxies have a younger stellar population in the central regions than their outskirts. 

(\textit{vi}) 58.3\% of the DPGs experienced external processes, characterized by tidal features, companion galaxies, as well as gas-star misalignments. This fraction is about twice as much as that of the control sample, suggesting the origin of double-peaked emission line profiles is associated with external processes.

Combining all the observational results listed above, we suggest that double-peaked emission line profiles in 35 out of 36 DPGs primarily originate from rotating discs.
The remaining one galaxy shows clear outflow features.
8 out of 35 DPGs show symmetric profiles that indicate undisturbed rotating discs, and the other 27 DPGs exhibit asymmetric profiles, suggesting dynamic disturbances in the rotating discs.

\section*{Acknowledgements}
YMC acknowledges support from the National Natural Science Foundation of China, NSFC grants 12333002, the China Manned Space Project with NO. CMS-CSST-2025-A08.

Funding for the Sloan Digital Sky Survey IV has been provided by the Alfred P. Sloan Foundation, the U.S. Department of Energy Office of Science, and the Participating Institutions. SDSS-IV acknowledges support and resources from the Center for High-Performance Computing at the University of Utah. The SDSS website is \href{https://www.sdss.org}{www.sdss.org.}

SDSS-IV is managed by the Astrophysical Research Consortium for the Participating Institutions of the SDSS Collaboration including the Brazilian Participation Group, the Carnegie Institution for Science, Carnegie Mellon University, the Chilean Participation Group, the French Participation Group, Harvard-Smithsonian Center for Astrophysics, Instituto de Astrofísica de Canarias, The Johns Hopkins University, Kavli Institute for the Physics and Mathematics of the Universe (IPMU) / University of Tokyo, Lawrence Berkeley National Laboratory, Leibniz Institut für Astrophysik Potsdam (AIP), Max-Planck-Institut für Astronomie (MPIA Heidelberg), Max-Planck-Institut für Astrophysik (MPA Garching), Max-Planck-Institut für Extraterrestrische Physik (MPE), National Astronomical Observatories of China, New Mexico State University, New York University, University of Notre Dame, Observatório Nacional / MCTI, The Ohio State University, Pennsylvania State University, Shanghai Astronomical Observatory, United Kingdom Participation Group, Universidad Nacional Autónoma de México, University of Arizona, University of Colorado Boulder, University of Oxford, University of Portsmouth, University of Utah, University of Virginia, University of Washington, University of Wisconsin, Vanderbilt University, and Yale University.
\section*{Data Availability}
The data underlying this article will be shared on reasonable request
to the corresponding author.
\bibliographystyle{mnras}
\bibliography{references}
\bsp	
\label{lastpage}
\end{document}